\documentclass[aps,prd,reprint,superscriptaddress,nofootinbib,longbibliography,amsmath,amssymb,floatfix]{revtex4-2}
\usepackage{graphicx}
\usepackage{bm}
\usepackage{booktabs}
\usepackage[above]{placeins}
\usepackage{hyperref}
\hypersetup{colorlinks=true,linkcolor=blue,citecolor=blue,urlcolor=blue,
pdftitle={Calibrated post-TOV parameter selection for joint inference of the symmetry-energy slope and gravity},
pdfauthor={Jiangshan Peng and Dao-Jun Liu}}
\newcommand{\lsym}{L_{\mathrm{sym}}}
\newcommand{\ms}{M_{\mathrm{s}}}
\newcommand{\nsat}{n_{\mathrm{sat}}}
\newcommand{\dr}{D_R}
\newcommand{\mstar}{\mu_1^{\ast}}
\newcommand{\rstar}{D_R^{\ast}}
\begin{document}
\title{Calibrated post-TOV parameter selection for joint inference\texorpdfstring{\\}{ }of the symmetry-energy slope and gravity}
\author{Jiangshan Peng}
\affiliation{Department of Physics, College of Mathematics and Sciences, Shanghai Normal University, Shanghai 200234, China}
\author{Dao-Jun Liu}
\email[Contact author: ]{djliu@shnu.edu.cn}
\affiliation{Department of Physics, College of Mathematics and Sciences, Shanghai Normal University, Shanghai 200234, China}

\begin{abstract}
Modifications of gravity and changes in the dense-matter equation of state (EOS) can produce similar neutron-star mass--radius relations, complicating EOS inference from observations. To quantify this compensation in a reduced gravity model, we compare nine post-Tolman--Oppenheimer--Volkoff corrections at a common strength relative to the corresponding general-relativistic (GR) term. Under the standard normalization, this comparison favors the mass-equation coefficient $\mu_1$ for all four benchmark EOSs. We retain $\mu_1$ and vary a single input to the nuclear energy functional, the symmetry-energy slope $\lsym$ at saturation. For shifts of $\pm10$ MeV from the $\lsym=40$ MeV GR reference, fitting one $\mu_1$ per sequence reduces the root-mean-square radius difference by about $64\%$ at fixed enclosed mass over $[1,2]M_\odot$. At fixed asymptotic mass, the reduction is $17\%$ for the lower slope and $75\%$ for the higher slope. The negative coefficient preferred at the lower slope and fixed enclosed mass yields a sequence that falls short of $2M_\odot$ in asymptotic mass. We perform joint Bayesian inference of $\lsym$ and $\mu_1$ using public posteriors obtained with GR measurement models for PSR~J0740+6620, PSR~J0437$-$4715, GW170817, and GW190425. For the two gravitational-wave events we relate compactness to tidal deformability through a GR fit. The central $95\%$ credible interval for $\mu_1$ is $[-0.09,0.97]$ and includes the GR value of zero. For this EOS family and crust--core matching, the four sources constrain $\lsym$ to nearly the same range in GR and in the single-$\mu_1$ model. The central $68\%$ intervals are $[28,128]$ and $[28,127]$ MeV, respectively, with medians of $73$ and $69$ MeV.
\end{abstract}

\maketitle

\section{Introduction}
\label{sec:intro}
Neutron stars probe the equation of state (EOS) of neutron-rich matter at densities beyond those accessible in terrestrial nuclei~\cite{LattimerPrakash2001}. In general relativity (GR), the Tolman--Oppenheimer--Volkoff (TOV) equations relate this EOS to stellar masses and radii~\cite{Tolman1939,OppenheimerVolkoff1939}. A key nuclear quantity is the symmetry-energy slope $\lsym$, which characterizes the density dependence of the energy cost of neutron--proton asymmetry at nuclear saturation. Its influence on the pressure of neutron-rich matter near saturation connects laboratory constraints on nuclear interactions to neutron-star radii~\cite{LattimerLim2013}. Inferring $\lsym$ from stellar observations therefore requires understanding both the matter model and the assumed gravitational dynamics.

Observations now constrain complementary aspects of neutron-star structure. X-ray pulse-profile modeling with the Neutron Star Interior Composition Explorer (NICER) provides joint mass--radius constraints, including those for PSR~J0740+6620 and PSR~J0437$-$4715~\cite{Salmi2024J0740,Miller2026J0437}. Gravitational-wave (GW) observations constrain the tidal response of neutron stars in binaries, as demonstrated by the EOS analysis of GW170817~\cite{Abbott2018GW170817EOS}. Combining these measurements with nuclear-theory calculations has tightened EOS constraints and allowed their dependence on high-density parametrizations to be assessed~\cite{Rutherford2024}. These analyses generally assume GR in both stellar structure and the interpretation of the signals. This dependence on GR motivates examining how the inferred matter properties change when gravitational dynamics are allowed to vary.

The resulting EOS--gravity degeneracy depends on the gravity theory and the observable under consideration. Cooney et al.~\cite{Cooney2010} showed that changes in mass--radius relations in perturbatively constrained $f(R)$ gravity can be difficult to distinguish from EOS effects. He et al.~\cite{He2015} examined this issue for the nuclear symmetry energy in scalar--tensor models constrained by binary-pulsar tests. The remaining gravity effects in those models were smaller than the symmetry-energy uncertainties. Thus, a degeneracy found in one model need not have the same importance in another.

Tidal observables extend this comparison beyond equilibrium masses and radii. Hinderer's calculation of relativistic neutron-star Love numbers~\cite{Hinderer2008} and the tidal-response formulations of Binnington and Poisson~\cite{BinningtonPoisson2009} and Damour and Nagar~\cite{DamourNagar2009} established a basis for these comparisons. Flanagan and Hinderer~\cite{FlanaganHinderer2008} showed how tidal effects in inspiral waveforms can constrain the EOS. Within GR, a study of compact objects with thin shells further illustrated the dependence of Love numbers on matter distribution~\cite{YangXieLiu2023}.

Beyond GR, studies in $R^2$ gravity~\cite{YazadjievDonevaKokkotas2018}, Rastall gravity~\cite{MengLiu2021}, and unimodular gravity with a nonconserved energy-momentum tensor~\cite{YangXieLiu2022} have examined how gravitational dynamics affect neutron-star tidal responses. Joint inference can assess these dependencies within a specified model. Biswas et al.~\cite{Biswas2024} developed a hierarchical Bayesian analysis of an EOS catalog and a coupling in a particular four-dimensional Horndeski theory.

The post-Tolman--Oppenheimer--Volkoff (post-TOV) formalism parametrizes deviations from the GR equations of stellar equilibrium. Glampedakis et al.~\cite{Glampedakis2015postTOV} classified corrections to the TOV equations by their radial profiles and their effects on stellar masses and radii. Their subsequent treatment of the exterior related post-TOV parameters to redshifts, apparent radii, and orbital frequencies~\cite{Glampedakis2016postTOVapps}. Silva and Yunes~\cite{SilvaYunes2019} used GW170817 to constrain a combination of these parameters through its effect on the GW phase. They also showed that different choices of EOS and gravity can occupy overlapping regions of the mass--radius plane. This leaves a quantitative question: over a specified mass interval, how much of the radius change caused by varying $\lsym$ can one gravity coefficient reproduce?

We first compare the nine second post-Newtonian corrections at a common radial root-mean-square (RMS) strength relative to the corresponding GR term. This comparison selects the mass-equation coefficient $\mu_1$ under the standard normalization. We then vary $\lsym$ in a nuclear metamodel and fit one value of $\mu_1$ along each mass--radius sequence. We perform the fit at fixed enclosed mass and at fixed asymptotic mass, which is the mass entering the observational analysis. Finally, we compare constraints on $\lsym$ with $\mu_1$ fixed in GR and free in post-TOV, using the same two NICER pulsars and two GW events. The comparison retains the GR measurement models underlying the public posteriors and uses a compactness--tidal-deformability relation for the GW events.

The remainder of this paper is organized as follows. In Sec.~\ref{sec:reduction}, we compare the nine post-TOV corrections and select  the most significant one. In Sec.~\ref{sec:degeneracy},  the EOS--gravity degeneracy is quantified through mass--radius comparisons. Then,  we combine NICER and GW data to constrain the two parameters in Sec.~\ref{sec:observations}, and finally discuss the results and their limitations in Sec.~\ref{sec:discussion}. Besides, some detailed numerical checks are given in Appendix~\ref{app:numerics}.

\section{Selection of a post-TOV coefficient}
\label{sec:reduction}
\subsection{Post-TOV equations and coefficients}
We compare corrections to the pressure and mass equations for static, spherically symmetric stars in geometrized units $G=c=1$. Solar System and binary-pulsar tests tightly constrain the 1PN departures from GR in the post-TOV formalism~\cite{Glampedakis2015postTOV}. We therefore set the corresponding coefficients to zero and retain the 2PN corrections. The stellar structure equations then read
\begin{align}
\frac{dp}{dr}&=F_p-\frac{\rho m}{r^2}\mathcal P_2,\\
\frac{dm}{dr}&=F_m+4\pi r^2\rho\mathcal M_2,
\label{eq:fullposttov}
\end{align}
where $r$ denotes the areal radius and $m(r)$ the interior mass function. The variables $p$, $\rho$, and $\varepsilon$ denote the pressure, rest-mass density, and total energy density, respectively.

The GR terms in the two equations are~\cite{OppenheimerVolkoff1939}
\begin{align*}
F_p&=-\frac{(\varepsilon+p)(m+4\pi r^3p)}{r^2(1-2m/r)},\\
F_m&=4\pi r^2\varepsilon.
\end{align*}
The dimensionless 2PN correction functions in the pressure and mass equations are~\cite{Glampedakis2015postTOV}
\begin{align}
\mathcal P_2&=\pi_1\frac{m^3}{r^5\rho}+\pi_2\frac{m^2}{r^2}
+\pi_3r^2p+\pi_4\frac{\Pi p}{\rho},\\
\mathcal M_2&=\mu_1\frac{m^3}{r^5\rho}+\mu_2\frac{m^2}{r^2}
+\mu_3r^2p+\mu_4\frac{\Pi p}{\rho}+\mu_5\frac{\Pi^3r}{m}.
\label{eq:secondorder}
\end{align}
Here $\Pi=(\varepsilon-\rho)/\rho$ is the specific internal energy. Four dimensionless coefficients, $\pi_1,\ldots,\pi_4$, enter the pressure equation, and five, $\mu_1,\ldots,\mu_5$, enter the mass equation. These coefficients parametrize departures from GR without specifying a particular theory of gravity~\cite{Glampedakis2015postTOV}. Clearly, setting all nine to zero recovers the TOV equations of GR.

\subsection{Measuring the relative strengths of the corrections}
Varying all nine coefficients would add nine free parameters to the EOS analysis. To study how modified gravity can mimic the effects of the nuclear symmetry energy on stellar radii, we seek a reduced model. To this end, we compare the contribution of each correction to the GR term in the same structure equation.

 We first compare the corrections on a fixed GR stellar profile. For each term, the coefficient magnitude required to reach a prescribed relative strength is
\begin{equation}
|a|=\eta\frac{\|F\|}{\|\Delta\|},
\label{eq:calibration}
\end{equation}
where the RMS value of a radial function $f(r)$ is defined by
\begin{equation}
\|f\|=\left[\frac{1}{R}\int_0^{R}f(r)^2\,dr\right]^{1/2}.
\label{eq:radialnorm}
\end{equation}
Here $a$ is the coefficient of a single correction, and $\Delta$ includes all factors multiplying $a$ in the structure equation. The reference term $F$ becomes $F_p$ for a pressure correction and $F_m$ for a mass correction. We prescribe the ratio $\eta>0$ of the RMS value of $a\Delta$ to that of $F$. The integral extends to the radius $R$ at the pressure cutoff.

To test how the comparison depends on the EOS, we use four tables from CompOSE~\cite{Typel2015CompOSE}. APR~\cite{Akmal1998}, BSK24~\cite{Goriely2013,Pearson2018,Pearson2019Erratum}, PT(GRDF2-DD2)~\cite{Typel2010,PaisTypel2017,Typel2018}, and NL3~\cite{Lalazissis1997,Xia2022,Xia2022Unified,Niu2026} span variational many-body, Skyrme, and relativistic mean-field descriptions of nuclear matter. NL3 provides a stiff EOS for comparison. The APR table combines the APR core with the Douchin--Haensel inner crust~\cite{Douchin2001} and the Baym--Pethick--Sutherland outer crust~\cite{Baym1971BPS}.

For each EOS, the GR TOV equations are integrated down to the lowest positive pressure in the table and the central pressure is adjusted so that the mass $M_{\rm cut}$ enclosed at $R$ is $1.4M_\odot$. We evaluate both RMS values in Eq.~\eqref{eq:calibration} on the same GR profile. With the coefficients defined as in the original post-TOV equations, a smaller $|a|$ at fixed $\eta$ means a larger relative correction per unit coefficient.

The post-TOV formalism was developed to describe small departures from GR~\cite{Glampedakis2015postTOV}. We choose $\eta\ll1$ so that the RMS value of each correction is small compared with that of its GR term. We take $\eta=10^{-3}$ as the reference value and repeat the calculation at $2.5\times10^{-4}$, $5\times10^{-4}$, $2\times10^{-3}$, and $4\times10^{-3}$. These factor-of-two steps allow us to test whether the ordering changes with $\eta$ when we recompute the stellar profiles.

An individual correction also changes the stellar profile on which its strength is evaluated. To include this effect, we solve the post-TOV equations with one nonzero coefficient at a time, treating the two signs separately. Furthermore, we use local continuation to vary the coefficient and central pressure $p_{\rm c}$, seeking a solution that satisfies Eq.~\eqref{eq:calibration} at $M_{\rm cut}=1.4M_\odot$. For each trial we reintegrate the structure equations and evaluate both RMS values up to the new cutoff radius using Eq.~\eqref{eq:radialnorm}.

For the post-TOV solutions we require a smooth center, with $m(r)=\mathcal O(r^3)$ and $dm/dr=\mathcal O(r^2)$ as $r\to0$. If $\rho_{\rm c}\Pi_{\rm c}^3\ne0$, a nonzero $\mu_5$ gives a contribution to $dm/dr$ that approaches a nonzero constant at the center. Here $\rho_{\rm c}$ and $\Pi_{\rm c}$ are the central values of $\rho$ and $\Pi$. The original post-TOV classification admits this term in family~$F_5$ under a weaker regularity condition~\cite{Glampedakis2015postTOV}. We therefore include $\mu_5$ in the comparison on fixed GR profiles and set it to zero when solving for modified profiles.

Finally, we solve for stellar profiles with the remaining eight corrections present together. We set $\pi_i=\mu_i=a$ for $i=1,\ldots,4$, keep $\mu_5=0$, and solve the post-TOV equations at the same enclosed mass for $a=\pm0.01,\pm0.05,\pm0.10$. On each resulting profile we use Eq.~\eqref{eq:radialnorm} to compute $Q_j$, the ratio of the RMS value of correction $j$ to that of its corresponding GR term. Here the coefficients are held equal, so the term with the largest $Q_j$ gives the largest relative correction.

\subsection{Comparison across four equations of state}
On the fixed GR profiles, $\mu_1$ requires the smallest coefficient magnitude for each of the four EOSs (TABLE~\ref{tab:calibration}). At $\eta=10^{-3}$, the next smallest is $|\pi_1|$, which is $2.67$--$3.09$ times as large as $|\mu_1|$. On a fixed profile, $F$ and $\Delta$ in Eq.~\eqref{eq:calibration} are independent of both $\eta$ and the sign of $a$. The required magnitudes are therefore the same for both signs and scale linearly with $\eta$, leaving the ordering unchanged at all five strengths.

\begin{table}[t]
\caption{Coefficient magnitudes $|a|$ required to reach $\eta=10^{-3}$ on fixed GR profiles at $M_{\rm cut}=1.4M_\odot$. Bold type marks the smallest magnitude for each EOS. DD2 denotes PT(GRDF2-DD2), and $R$ is the radius at the lowest positive pressure in each EOS table.}
\label{tab:calibration}
\setlength{\tabcolsep}{4pt}
\begin{ruledtabular}
\begin{tabular}{lrrrr}
Coefficient & APR & BSK24 & DD2 & NL3\\
$\pi_1$ & 0.01551 & 0.01794 & 0.01879 & 0.02205\\
$\pi_2$ & 0.1065 & 0.1208 & 0.1266 & 0.1507\\
$\pi_3$ & 1.058 & 1.303 & 1.406 & 1.699\\
$\pi_4$ & 0.2408 & 0.3449 & 0.3933 & 0.4692\\
$\mu_1$ & \textbf{0.005026} & \textbf{0.006347} & \textbf{0.006863} & \textbf{0.008251}\\
$\mu_2$ & 0.04194 & 0.05127 & 0.05602 & 0.06887\\
$\mu_3$ & 0.6339 & 0.8461 & 0.9351 & 1.198\\
$\mu_4$ & 0.2262 & 0.3600 & 0.4119 & 0.5179\\
$\mu_5$ & 0.2136 & 0.3833 & 0.4783 & 0.5748\\
\hline
$R$ (km) & 11.338 & 12.574 & 13.172 & 14.581\\
\end{tabular}
\end{ruledtabular}
\end{table}

\begin{table*}[t]
\caption{Coefficient magnitudes $|a|$ required to reach $\eta=10^{-3}$ when the post-TOV equations are solved with one nonzero coefficient at $M_{\rm cut}=1.4M_\odot$. The $+$ and $-$ columns give the two coefficient signs. Bold type marks the smallest available magnitude in each column. Dashes indicate cases excluded by central regularity ($\mu_5\ne0$) or cases for which no solution was obtained ($\pi_1<0$). The profiles end at the EOS pressure cutoff. Their dynamical stability has not been tested.}
\label{tab:selfconsistent-amplitudes}
\begin{ruledtabular}
\begin{tabular}{l*{8}{r}}
Coefficient & \multicolumn{2}{c}{APR} & \multicolumn{2}{c}{BSK24} & \multicolumn{2}{c}{DD2} & \multicolumn{2}{c}{NL3}\\
 & $+$ & $-$ & $+$ & $-$ & $+$ & $-$ & $+$ & $-$\\
\hline
$\pi_{1}$ & $0.01583$ & \textemdash & $0.01840$ & \textemdash & $0.01926$ & \textemdash & $0.02259$ & \textemdash\\
$\pi_{2}$ & $0.1065$ & $0.1065$ & $0.1208$ & $0.1208$ & $0.1266$ & $0.1267$ & $0.1506$ & $0.1508$\\
$\pi_{3}$ & $1.057$ & $1.059$ & $1.301$ & $1.304$ & $1.405$ & $1.407$ & $1.698$ & $1.701$\\
$\pi_{4}$ & $0.2405$ & $0.2412$ & $0.3444$ & $0.3454$ & $0.3927$ & $0.3939$ & $0.4686$ & $0.4699$\\
$\mu_{1}$ & $\mathbf{0.005025}$ & $\mathbf{0.005028}$ & $\mathbf{0.006346}$ & $\mathbf{0.006349}$ & $\mathbf{0.006861}$ & $\mathbf{0.006865}$ & $\mathbf{0.008249}$ & $\mathbf{0.008254}$\\
$\mu_{2}$ & $0.04193$ & $0.04194$ & $0.05127$ & $0.05127$ & $0.05602$ & $0.05602$ & $0.06886$ & $0.06887$\\
$\mu_{3}$ & $0.6336$ & $0.6341$ & $0.8458$ & $0.8464$ & $0.9348$ & $0.9354$ & $1.198$ & $1.198$\\
$\mu_{4}$ & $0.2255$ & $0.2269$ & $0.3588$ & $0.3611$ & $0.4107$ & $0.4131$ & $0.5165$ & $0.5194$\\
$\mu_{5}$ & \textemdash & \textemdash & \textemdash & \textemdash & \textemdash & \textemdash & \textemdash & \textemdash\\
\end{tabular}
\end{ruledtabular}
\end{table*}

\begin{table*}[t]
\caption{Relative strengths $10^3Q_j$ for eight equal coefficients $\pi_i=\mu_i=a$ ($i=1,\ldots,4$), with $\mu_5=0$, $M_{\rm cut}=1.4M_\odot$, and $a=\pm0.01$. Each $Q_j$ is the ratio of the RMS value of correction $j$ to that of its GR term on the same stellar profile. The $+$ and $-$ columns give the sign of $a$. Bold type marks the largest strength in each column. Dashes indicate that no DD2 solution was obtained for $a<0$. For APR at $a=-0.01$, the finite-pressure cutoff is at $R\simeq1553$ km.}
\label{tab:equal-amplitude-strengths}
\begin{ruledtabular}
\begin{tabular}{l*{8}{r}}
Term $j$ & \multicolumn{2}{c}{APR} & \multicolumn{2}{c}{BSK24} & \multicolumn{2}{c}{DD2} & \multicolumn{2}{c}{NL3}\\
 & $+$ & $-$ & $+$ & $-$ & $+$ & $-$ & $+$ & $-$\\
\hline
$\pi_{1}$ & $0.6339$ & $0.6658$ & $0.5458$ & $0.5777$ & $0.5216$ & \textemdash & $0.4448$ & $0.4683$\\
$\pi_{2}$ & $0.09394$ & $0.09388$ & $0.08282$ & $0.08278$ & $0.07900$ & \textemdash & $0.06638$ & $0.06634$\\
$\pi_{3}$ & $0.009454$ & $0.009457$ & $0.007679$ & $0.007679$ & $0.007114$ & \textemdash & $0.005886$ & $0.005884$\\
$\pi_{4}$ & $0.04156$ & $0.04153$ & $0.02902$ & $0.02899$ & $0.02545$ & \textemdash & $0.02133$ & $0.02130$\\
$\mu_{1}$ & $\mathbf{1.890}$ & $\mathbf{2.487}$ & $\mathbf{1.486}$ & $\mathbf{1.987}$ & $\mathbf{1.375}$ & \textemdash & $\mathbf{1.144}$ & $\mathbf{1.516}$\\
$\mu_{2}$ & $0.2386$ & $0.2386$ & $0.1951$ & $0.1951$ & $0.1786$ & \textemdash & $0.1452$ & $0.1452$\\
$\mu_{3}$ & $0.01579$ & $0.01578$ & $0.01183$ & $0.01182$ & $0.01070$ & \textemdash & $0.008352$ & $0.008345$\\
$\mu_{4}$ & $0.04430$ & $0.04420$ & $0.02783$ & $0.02776$ & $0.02431$ & \textemdash & $0.01933$ & $0.01929$\\
\end{tabular}
\end{ruledtabular}
\end{table*}

For the single-coefficient calculations, four EOSs, nine coefficients, two signs, and five target strengths give $360$ combinations. The central regularity condition excludes the $40$ cases with nonzero $\mu_5$. We obtain $300$ solutions that satisfy both Eq.~\eqref{eq:calibration} and the enclosed-mass constraint. In the remaining $20$ cases, all with $\pi_1<0$, integration stopped before the pressure cutoff because a trial step failed the checks for finite derivatives and outwardly decreasing pressure. A negative $\pi_1$ contributes positively to $dp/dr$ and opposes the GR pressure gradient. Glampedakis et al.~\cite{Glampedakis2015postTOV} also reported difficulty locating the surface for this sign as the pressure gradient approached zero in the outer layers. Whether solutions meeting both constraints exist in these cases remains open.

TABLE~\ref{tab:selfconsistent-amplitudes} gives the coefficient magnitudes on the modified profiles at $\eta=10^{-3}$. Among the computed solutions, $|\mu_1|$ is the smallest for every EOS, sign, and target strength. The next smallest is $|\pi_1|$ for positive coefficients and $|\mu_2|$ among the available solutions for negative coefficients. Recomputing the profiles changes $|\mu_1|$ from its fixed-GR value by at most $0.0291\%$ at $\eta=10^{-3}$ and $0.117\%$ over all five strengths. The accompanying data give the results at every strength.

With eight equal coefficients, we obtain solutions for $21$ of the $24$ combinations of EOS and coefficient value. For DD2 at the three negative values of $a$, the integration could not follow the branch across the low-pressure density jump while maintaining an outwardly decreasing pressure. The outcome may depend on the treatment of the interface or the choice of equilibrium branch.

In all $21$ solutions, $Q_{\mu_1}$ is the largest relative strength and $Q_{\pi_1}$ the second largest. Their ratio ranges from $2.53$ to $3.74$. The strengths themselves vary with the EOS. For example, at $a=+0.01$, $Q_{\mu_1}$ is $1.890 \times 10^{-3}$ for APR and $1.144 \times 10^{-3}$ for NL3 (see TABLE~\ref{tab:equal-amplitude-strengths}).

Independent integrations of the $300$ single-coefficient solutions reproduce their masses, radii, and RMS ratios with relative differences below $10^{-6}$ (see Appendix~\ref{app:calibrationchecks} for the details).

\subsection{Choice of \texorpdfstring{$\mu_1$}{mu1} for the EOS analysis}
All four EOSs lead to the same choice at $M_{\rm cut}=1.4M_\odot$ with the original post-TOV coefficient definitions. Among the solutions obtained, $\mu_1$ requires the smallest magnitude to reach a given relative strength, both on fixed GR profiles and after the stellar structure is recomputed. When the eight coefficients have the same value, the $\mu_1$ term gives the largest relative correction (see Tables~\ref{tab:calibration}--\ref{tab:equal-amplitude-strengths}). We therefore vary only $\mu_1$ in the subsequent EOS analysis and set the other post-TOV coefficients to zero.

This choice reflects the size of the corrections in the stellar structure equations. At equal RMS strength in the reference EOS considered in Sec.~\ref{sec:eosfamily}, the $\mu_2$ term can produce a larger radius response than $\mu_1$ (see Appendix~\ref{app:calibrationchecks}). It should be noted that, we have not tested which coefficient observations would impose the tightest constraints, nor have we quantified the effect of omitting the other terms. The coefficient values required to reach a prescribed level $\eta$ do not establish the priors for inference or define the range over which the reduced model is valid.

A small RMS ratio does not guarantee a small correction at every radius. In the outer layers of some of the models used for these comparisons, the enclosed mass $m(r)$ decreases with increasing radius. Some equal-coefficient solutions with $a<0$ have extended envelopes (see TABLE~\ref{tab:equal-amplitude-strengths}). For $\pi_1\ne0$ we stop at finite pressure. Smooth matching to the vacuum exterior of Ref.~\cite{Glampedakis2016postTOVapps} requires $\pi_1=0$ because its contribution to the metric derivative diverges at a zero-density surface. So we use these profiles only to compare the terms in the structure equations. For the radius comparisons and observational inference, we compute separate sequences with only $\mu_1$ free.

\section{EOS--gravity degeneracy in mass--radius relations}
\label{sec:degeneracy}
In Sec.~\ref{sec:reduction}, we identified $\mu_1$ as the leading term by comparing corrections to the stellar structure equations. 
A natural question then arises: how well can a change in $\mu_1$ reproduce the radius changes caused by varying the symmetry-energy slope $\lsym$?
To address this question, we will compare stars with the same asymptotic mass—specifically, the mass used in the observational analysis—and repeat the fits at the same enclosed mass to isolate the effect of exterior mass conversion.

\subsection{Stellar equations with \texorpdfstring{$\mu_1$}{mu1}}
\label{sec:model}
With $\mu_1$ as the only free gravity coefficient, the pressure and enclosed mass satisfy
\begin{align}
\frac{dp}{dr}&=-\frac{(\varepsilon+p)(m+4\pi r^3p)}{r^2(1-2m/r)},
\label{eq:pressure}\\
\frac{dm}{dr}&=4\pi r^2\varepsilon+4\pi\mu_1\frac{m^3}{r^3}.
\label{eq:mass}
\end{align}
The EOS supplies $\varepsilon(p)$. The $\mu_1$ term changes the enclosed mass directly and thereby changes the pressure gradient in Eq.~\eqref{eq:pressure}~\cite{Glampedakis2015postTOV}.

We impose the regular central conditions $p(0)=p_{\rm c}$ and $m(0)=0$. The $\mu_1$ correction enters $dm/dr$ at order $r^6$, so the usual central expansion $m(r)=(4\pi/3)\varepsilon_{\rm c}r^3+\mathcal O(r^5)$ remains valid. Here $\varepsilon_{\rm c}$ is the central energy density.

\subsection{EOS family with varying \texorpdfstring{$\lsym$}{Lsym}}
\label{sec:eosfamily}
The four EOSs used to select $\mu_1$ differ in several nuclear properties. To isolate the effect of $\lsym$, we now vary it within a single nuclear metamodel~\cite{Margueron2018a,Margueron2018b} with a sound-speed extension~\cite{Tews2018}. Our implementation is adapted from the public JESTER code~\cite{Wouters2025}. The slope that we vary in the nuclear energy functional is
\begin{equation}
\lsym=3\nsat\left.\frac{dS}{dn}\right|_{\nsat},\;
S(n)=e_{\rm PNM}(n)-e_{\rm SNM}(n),
\label{eq:symmetrydefinition}
\end{equation}
where $n$ is the baryon number density, $\nsat$ is the saturation density, and $e_{\rm PNM}$ and $e_{\rm SNM}$ are the nuclear energies per baryon in pure neutron matter and symmetric matter. The difference defines the symmetry energy used for the metamodel inputs. The distinction from the quadratic coefficient in isospin asymmetry is given in Appendix~\ref{app:eosimplementation}.

\begin{table}[tb]
\caption{EOS inputs held fixed in the mass--radius comparisons. Nuclear energy coefficients are in MeV.}
\label{tab:eos}
\begin{ruledtabular}
\begin{tabular}{ll}
Input & Value or prescription\\
$\nsat$ & $0.16~\mathrm{fm}^{-3}$\\
$(E_{\rm sat},K_{\rm sat},Q_{\rm sat},Z_{\rm sat})$ & $(-16,200,0,0)$\\
$(E_{\rm sym},K_{\rm sym},Q_{\rm sym},Z_{\rm sym})$ & $(32,-100,0,0)$\\
Effective-mass corrections & All six coefficients zero\\
Independent potential $\delta^4$ terms & Zero\\
Last retained crust density & $0.07596~\mathrm{fm}^{-3}$\\
Metamodel starts at & $0.75\nsat$\\
Sound-speed extension starts at & $1.5\nsat$\\
Extension nodes & $n/\nsat=3,4,\ldots,10$\\
Maximum tabulated density & $12\nsat$\\
$c_s^2$ at nodes and endpoint & $0.5$\\
\end{tabular}
\end{ruledtabular}
\end{table}

We determine the equilibrium composition of charge-neutral neutron--proton--electron matter by minimizing the full energy density, and evaluate $c_s^2=dp/d\varepsilon$ along this equilibrium curve. We retain the low-density sampling of the Douchin--Haensel crust~\cite{Douchin2001} and connect its sound speed to the metamodel at $0.75\nsat$ via cubic Hermite interpolation. Starting from the pressure and baryon chemical potential in the crust, we integrate the thermodynamic relations to obtain the pressure used in the stellar structure equations. At a given density, this pressure thus depends on both the metamodel input $\lsym$ and the crust--core matching prescription (see Appendix~\ref{app:eosimplementation}).

From $1.5\nsat$ to $3\nsat$, we interpolate $c_s^2$ linearly to its prescribed value of $0.5$. The constant sound speed above $n_a=3\nsat$ then gives
\begin{equation}
p-p_a=\tfrac12(\varepsilon-\varepsilon_a).
\label{eq:highdensity}
\end{equation}
The matching pressure $p_a$ and energy density $\varepsilon_a$ vary with $\lsym$ even with all inputs in TABLE~\ref{tab:eos} fixed. The high-density EOS therefore contributes to the radius response.

Every EOS shares the same low-density crust. We integrate to its lowest tabulated pressure, $p_{\rm cut}=1.0611\times10^{-19}~\mathrm{MeV\,fm}^{-3}$, and denote the radius and enclosed mass there by $R$ and $\ms=m(R)$. Matching to the post-TOV exterior at $R$ gives the asymptotic mass~\cite{Glampedakis2016postTOVapps}
\begin{equation}
M=\frac{\ms}{\sqrt{1-4\pi\mu_1(\ms/R)^2}}.
\label{eq:admmass}
\end{equation}
We require a positive radicand when using this relation. In GR, $M=\ms$. For $\mu_1<0$, the asymptotic mass is smaller than $\ms$, whereas for $\mu_1>0$ it is larger. Varying $\mu_1$ can thus change which points on a sequence are compared even when the mass interval is held fixed.

\begin{figure*}[t]
\includegraphics[width=\textwidth]{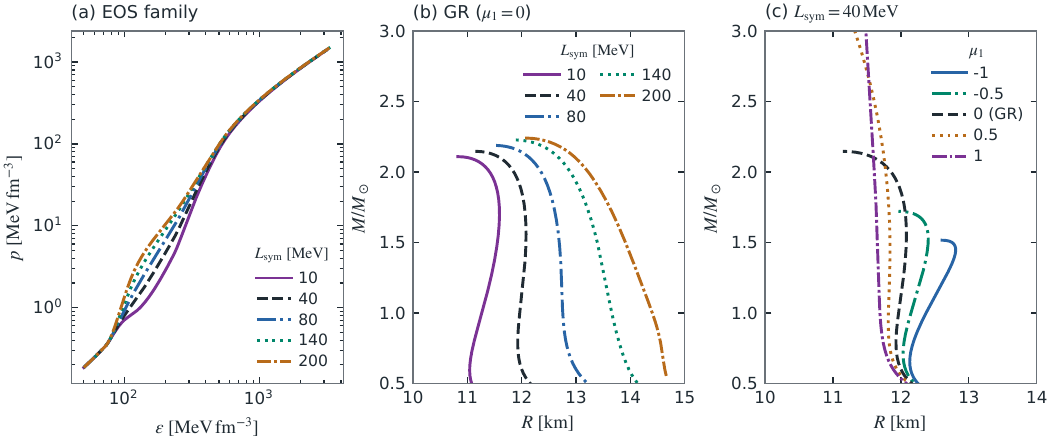}
\caption{Effects of varying $\lsym$ or $\mu_1$ separately. (a) Pressure versus energy density for selected EOSs. (b) Their mass--radius sequences in GR. (c) The effect of $\mu_1$ at $\lsym=40$ MeV. In (b) and (c) the vertical coordinate is the asymptotic mass $M$. 
}
\label{fig:eos}
\end{figure*}

\subsection{Radius comparison over a common mass interval}
\label{sec:methods}
To compare a candidate EOS with a GR reference, we construct stellar sequences over
\begin{equation}
\lsym\in[10,200]~\mathrm{MeV},\qquad \mu_1\in[-1,1].
\label{eq:box}
\end{equation}
For each parameter pair we vary the central pressure and evaluate $M$ from Eq.~\eqref{eq:admmass}. We select the first connected branch that satisfies the exterior matching condition and reaches $1M_\odot$ as $M$ increases. We require this branch to cover
\begin{equation}
M\in[1,2]M_\odot.
\label{eq:window}
\end{equation}
For the enclosed-mass comparison, we select the unique continuous branch increasing in $\ms$ that contains $1M_\odot$ and covers the same interval in $\ms$. Using the same endpoints isolates the effect of the mass definition. The branch selection gives a single radius at each mass. Dynamical stability would require a separate calculation.

Within each selected branch, we interpolate the radii at the comparison masses with a shape-preserving piecewise cubic Hermite interpolant~\cite{FritschButland1984}. We quantify the separation from a reference GR sequence at $\lsym=L_0$ by the RMS radius distance
\begin{equation}
\dr=\sqrt{\frac{1}{N_M}\sum_{k=1}^{N_M}(R_k-R_{0,k})^2},
\label{eq:distance}
\end{equation}
where $R_k$ and $R_{0,k}$ are the candidate and reference radii at the same mass, with $N_M=161$ equally spaced points including both endpoints. Either $M$ or $\ms$ is used  throughout a given sum.

For each candidate EOS, we first set $\mu_1=0$ to obtain $D_R^{\rm GR}$,  then fit one constant $\mu_1$ for the whole sequence and denote the smallest distance found by $\rstar$ and its coefficient by $\mstar$. Candidate minima and both edges of the sampled mass-coverage intervals are refined (see Appendix~\ref{sec:verification} for the details). On the $39\times21$ display grid, $482$ of the $819$ sequences cover the window in $M$ and $767$ cover it in $\ms$.

\begin{figure*}[t]
\centering
\includegraphics[width=0.96\textwidth]{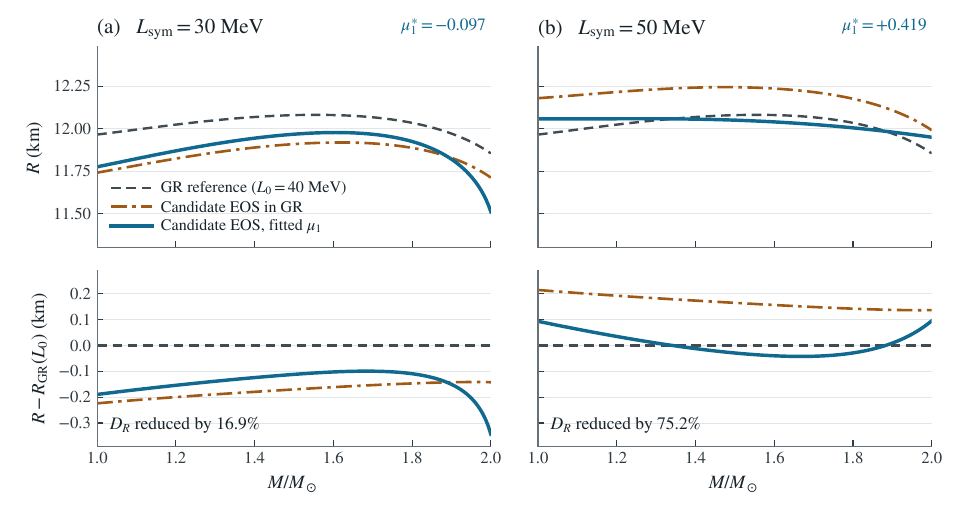}
\caption{Near-reference radius fits at fixed asymptotic mass $M$. The upper panels compare the $\lsym=30$ and $50$ MeV sequences with the $L_0=40$ MeV GR reference before and after fitting $\mu_1$. The lower panels show the signed radius difference at each mass. Percentages give the reduction in the RMS distance over $[1,2]M_\odot$ (Table~\ref{tab:results}).}
\label{fig:nearreference}
\end{figure*}

\begin{table}[tbp]
\caption{Radius fits relative to $L_0=40$ MeV. Distances are in km. The GR distance is the same in the two mass coordinates. Each fitted $\mstar$ is held constant along its sequence.}
\label{tab:results}
\begin{ruledtabular}
\begin{tabular}{rrrrrr}
 & & \multicolumn{2}{c}{Fixed $\ms$} & \multicolumn{2}{c}{Fixed $M$}\\
$\lsym$ (MeV) & $D_R^{\rm GR}$ & $\mstar$ & $\rstar$ & $\mstar$ & $\rstar$\\
10 & 0.544 & -0.529 & 0.289 & -0.087 & 0.513 \\
20 & 0.357 & -0.581 & 0.129 & -0.096 & 0.325 \\
30 & 0.176 & -0.295 & 0.064 & -0.097 & 0.146 \\
40 & 0.000 & 0.000 & 0.000 & 0.000 & 0.000 \\
50 & 0.169 & 0.308 & 0.060 & 0.419 & 0.042 \\
60 & 0.330 & 0.634 & 0.117 & 0.978 & 0.071 \\
65 & 0.407 & 0.806 & 0.143 & 1.000 & 0.118 \\
70 & 0.483 & 0.984 & 0.168 & 1.000 & 0.193 \\
80 & 0.628 & 1.000 & 0.255 & 1.000 & 0.351 \\
100 & 0.894 & 1.000 & 0.499 & 1.000 & 0.648 \\
140 & 1.341 & 1.000 & 0.953 & 1.000 & 1.157 \\
200 & 1.880 & 1.000 & 1.521 & 1.000 & 1.795 \\
\end{tabular}
\end{ruledtabular}
\end{table}

\subsection{Radius responses to \texorpdfstring{$\lsym$ and $\mu_1$}{Lsym and mu1}}
\label{sec:results}
In GR, increasing $\lsym$ from $30$ to $40$ and $50$ MeV increases the radius of a $1.4M_\odot$ star from $11.889$ to $12.068$ and $12.241$ km, respectively (see FIG.~\ref{fig:eos}). At fixed $\lsym=40$ MeV and the same asymptotic mass, increasing $\mu_1$ from zero to $0.5$ reduces the radius to $11.840$ km. The two parameters therefore shift the radius in opposite directions near the reference. For small parameter changes at fixed $M$,
\begin{equation}
\Delta R(M)\simeq
\left.\frac{\partial R}{\partial\lsym}\right|_{M,0}\Delta\lsym
+\left.\frac{\partial R}{\partial\mu_1}\right|_{M,0}\mu_1,
\label{eq:localresponse}
\end{equation}
where $\Delta\lsym=\lsym-L_0$, and the derivatives are evaluated at $(L_0,0)$. A coefficient that cancels the radius difference at one mass can cancel it throughout Eq.~\eqref{eq:window} only if the two derivatives have proportional mass dependence to this order.

\subsection{Radius compensation at fixed enclosed and asymptotic mass}
We first compare the two mass definitions near the GR reference at $L_0=40$ MeV. At fixed $\ms$, fitting $\mu_1$ reduces the RMS radius distance by about $64\%$ for both the $30$ and $50$ MeV candidates (see TABLE~\ref{tab:results}). Converting these fitted sequences to $M$ changes both the radii compared at a given mass and the mass range available for the fit.

For $\lsym=30$ MeV, the sequence fitted at fixed $\ms$ has $\mstar=-0.295$ and reaches only $M=1.840M_\odot$. At fixed $M$, the branch first covers the full interval near $\mu_1=-0.1134$. Refining the search on the allowed side gives a smaller distance at $\mstar=-0.097$, where the branch reaches $2.018M_\odot$. The RMS distance decreases from $0.1756$ to $0.1460$ km, a reduction of $16.9\%$ (see FIG.~\ref{fig:nearreference}). Thus the mass-coverage requirement excludes the more negative coefficient preferred at fixed $\ms$, although the fixed-$M$ minimum lies inside the allowed interval.

\begin{figure}[tbp]
\includegraphics[width=\columnwidth]{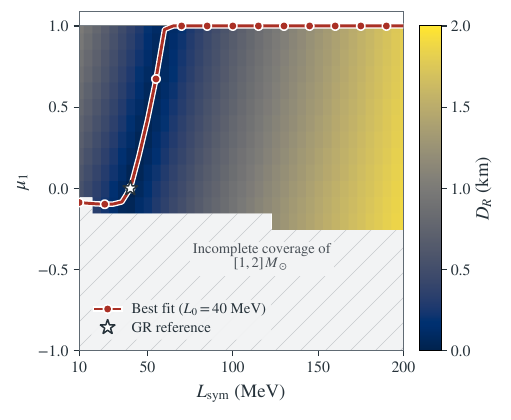}
\caption{RMS radius distance from the $L_0=40$ MeV GR reference (star) at fixed $M$. Colors give the computed distances. Hatching marks grid points whose sequences do not cover $[1,2]M_\odot$. The red curve connects independently refined fits and can pass through a cell whose center lacks coverage.}
\label{fig:map}
\end{figure}

For $\lsym=50$ MeV, the sequence fitted at fixed $\ms$ covers the entire interval in $M$. Its distance changes from $0.0603$ km at fixed $\ms$ to $0.0590$ km at fixed $M$. Refitting $\mu_1$ at fixed $M$ reduces the distance to $0.0418$ km, $75.2\%$ below the GR value. A shift of $\lsym$ by $+10$ MeV can therefore be compensated much more closely at the same asymptotic masses than a shift by $-10$ MeV. Both fits leave radius differences that vary with mass.

Farther above the reference, the fitted $\mu_1$ reaches its imposed bound of $1$. On the $5$ MeV grid this first occurs at $\lsym=65$ MeV for fixed $M$ (see FIG.~\ref{fig:map}). At $200$ MeV the remaining distance is $1.795$ km, only $4.5\%$ below the GR value. In FIG.~\ref{fig:examples} we show how these residuals vary with mass for three representative EOSs.

\subsection{Dependence on the reference EOS}
\begin{figure}[tbp]
\includegraphics[width=\columnwidth]{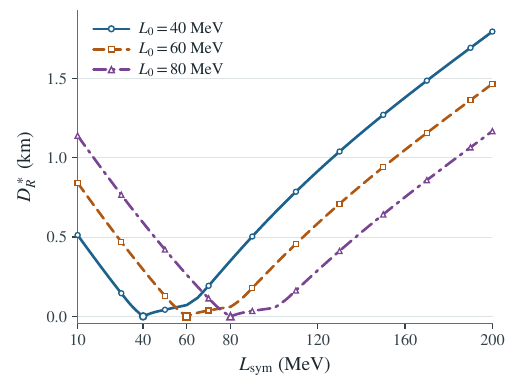}
\caption{Smallest computed RMS radius distance at fixed $M$ for GR references with $L_0=40$, $60$, and $80$ MeV. Each curve passes through 39 fits evaluated on the same 161 masses. Symbols mark selected fits. The curves use shape-preserving interpolation on either side of the exact zero at $\lsym=L_0$ and straight segments across changes in the active constraints.}
\label{fig:references}
\end{figure}

The mass-coverage and coefficient limits also depend on which GR radii are used as the reference. We repeat the fits at fixed $M$ for $L_0=60$ and $80$ MeV over the same mass and coefficient ranges. For $L_0=40$, $60$, and $80$ MeV, the fits first reach $\mu_1=1$ at the sampled slopes $65$, $80$, and $100$ MeV, respectively (See FIG.~\ref{fig:references}). Each curve gives the closest match to a prescribed EOS. To infer which parameter values are supported by data, we next compare the predicted radii and tidal deformabilities with observations.

\section{Joint NICER and GW constraints on \texorpdfstring{$\lsym$ and $\mu_1$}{Lsym and mu1}}
\label{sec:observations}
In Sec.~\ref{sec:degeneracy}, we show that the degree of radius compensation depends on the mass held fixed. Observations use the asymptotic mass $M$. We now use the sequences expressed in $M$ to show how much the inferred $\lsym$ changes when $\mu_1$ is free.

\subsection{Mass conversion and GR measurement assumptions}
We use the asymptotic mass of Eq.~\eqref{eq:admmass} for both the NICER and GW data. The mass conversion requires a positive radicand and specifies only the mass of the adopted exterior. For each parameter pair $\theta=(\lsym,\mu_1)$, we interpolate $R(M;\theta)$ along the first connected mass-increasing branch that reaches $1M_\odot$. For the likelihood we use each branch over its computed mass range, including those that fall short of $2M_\odot$. A rising mass along the sequence alone does not establish radial stability.

The inferred constraints are conditional on the GR measurement models used to obtain the public posteriors. For NICER we retain the original models of photon propagation, the rotating stellar surface, and the hotspots. We use the static circumferential radius of our model to approximate the reported equatorial radius. For the GW events the original GR waveforms are retained.

\subsection{Public posteriors and the GW tidal proxy}
\begin{table*}[t]
\caption{Public posterior samples used in the inference. The Zenodo records identify the data releases. For both GW events we use source-frame component masses, taking the released source-frame values where available.}
\label{tab:observations}
\begin{ruledtabular}
\begin{tabular}{lp{0.49\textwidth}rr}
Source & Selected analysis & Zenodo record & Rows\\
PSR J0740+6620 & \raggedright NICER and XMM-Newton, ST-U~\cite{Salmi2024J0740,Salmi2024J0740Data} & 10519473 & $1\,137\,394$\\
PSR J0437$-$4715 & \raggedright Preferred three-hotspot model with modulated nonthermal emission~\cite{Miller2026J0437,Miller2026J0437Data} & 17833896 & $2\,972\,029$\\
GW170817 & \raggedright IMRPhenomXP\_NRTidalv3, full frequency range with calibration uncertainty~\cite{Wang2026GW170817,WangGWPosteriorData} & 18489906 & $5\,863$\\
GW190425 & \raggedright GWTC-2.1 (release 2), low-spin IMRPhenomPv2\_NRTidal~\cite{Abbott2020GW190425,Abbott2024GWTC21,GW190425Data2022} & 6513631 & $20\,000$\\
\end{tabular}
\end{ruledtabular}
\end{table*}
In TABLE~\ref{tab:observations}, the selected sources for analysis are listed. For the two pulsars (NICER), we use weighted joint mass--radius samples, which already include the radio timing information. For the two binary mergers (GW), we take both component masses and both tidal deformabilities from the same posterior sample to preserve their correlations.

The single-$\mu_1$ model does not supply consistent tidal perturbation equations. To connect its mass--radius predictions to the GW tidal samples, we use the GR compactness--Love fit~\cite{YagiYunes2017Universal}
\begin{equation}
C_{\rm fit}(\Lambda)=0.360-0.0355\ln\Lambda
+0.000705(\ln\Lambda)^2,
\label{eq:clove}
\end{equation}
where $\Lambda$ is the dimensionless quadrupolar tidal deformability and $C_{\rm fit}(\Lambda)$ is the fitted compactness. The reported maximum relative deviation on the GR calibration sequences is approximately $6.5\%$. Using this value to set the error scale, we assume that at fixed stellar compactness $C=GM/(Rc^2)$ the fractional residual $[C_{\rm fit}(\Lambda)-C]/C$ is uniform on $[-0.065,0.065]$. The corresponding density in $\Lambda$ is
\begin{equation}
p_{\rm prox}(\Lambda\mid M,R)=
\frac{|dC_{\rm fit}/d\Lambda|}{0.13C}\,
\mathbf{1}_{[0.935C,\,1.065C]}(C_{\rm fit}),
\label{eq:proxy}
\end{equation}
on the positive, monotonically decreasing branch of the fit. The indicator $\mathbf{1}$ restricts $C_{\rm fit}$ to the stated interval of width $0.13C$, and the derivative is the Jacobian of the transformation. Our baseline analysis assumes independent compactness residuals for the two stars. Direct GR tidal solutions for $25$ stars in our EOS family give compactness residuals between $-1.93\%$ and $+3.09\%$ (see Appendix~\ref{app:finiteamplitude}). This checks the chosen width for those GR stars. For $\mu_1\ne0$ the relation remains an assumption.

\begin{figure}[t]
\centering
\includegraphics[width=\columnwidth]{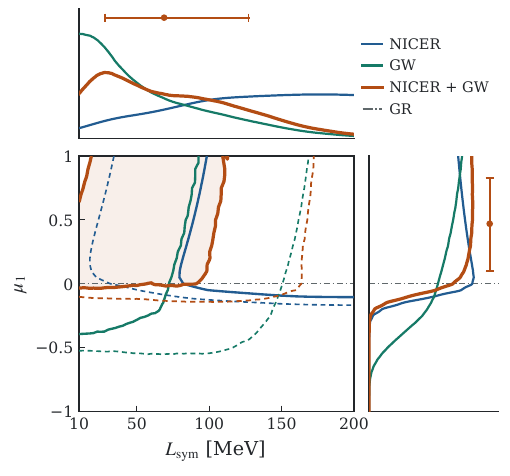}
\caption{Joint and marginal posteriors for NICER (blue), GW (green), and NICER+GW (orange). Solid and dashed contours enclose the $68\%$ and $95\%$ highest-posterior-density regions, respectively. Shading marks the joint $68\%$ region for NICER+GW. The marginal density curves above and at right pass through the grid values. Orange dots and bars give the corresponding four-source medians and central $68\%$ intervals from the inference grid. The gray dash-dotted line marks $\mu_1=0$.}
\label{fig:observationcorner}
\end{figure}

\begin{figure}[tb]
\includegraphics[width=\columnwidth]{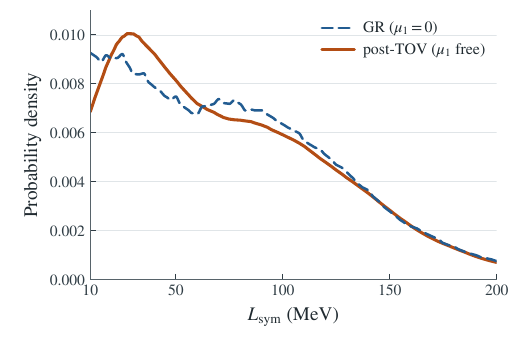}
\caption{Four-source marginal posterior density of $\lsym$ in $\mathrm{MeV}^{-1}$. The dashed blue curve fixes $\mu_1=0$ (GR), and the solid orange curve marginalizes over $\mu_1$ (post-TOV). As in FIG.~\ref{fig:observationcorner}, the marginal curves preserve the posterior probability in each grid interval and integrate to unity.}
\label{fig:observationL}
\end{figure}

\begin{figure*}[t]
\includegraphics[width=\textwidth]{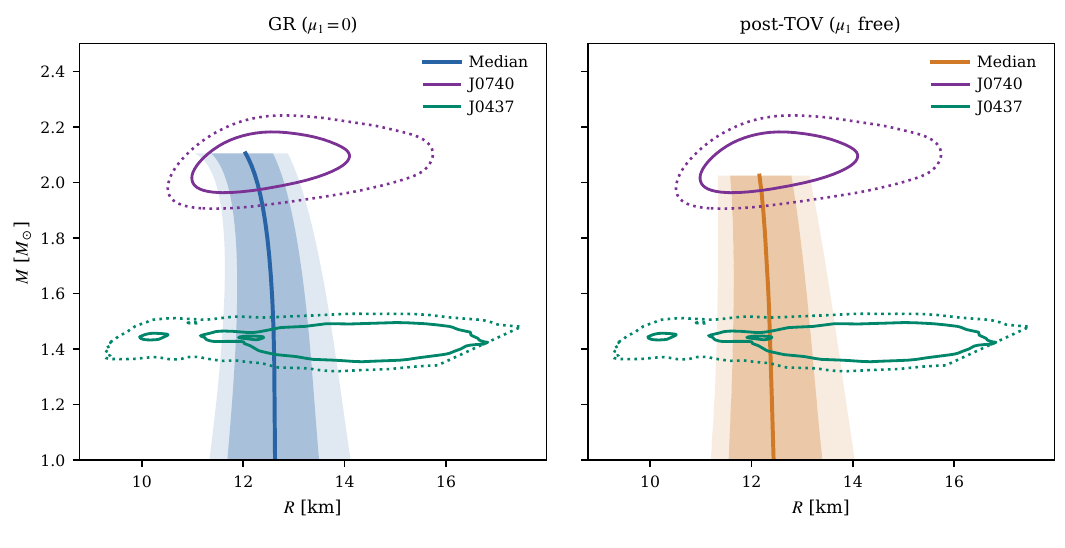}
\caption{Mass--radius predictions for GR (blue) and post-TOV (orange) from the same NICER+GW data. The mass coordinate is the asymptotic mass $M$. At each mass, the curve gives the posterior median radius and the dark and light bands give the central $68\%$ and $95\%$ radius intervals. We plot masses supported by at least $99\%$ of the original posterior weight. This sets the upper display limits of approximately $2.105M_\odot$ (GR) and $2.025M_\odot$ (post-TOV). Purple solid and dotted curves show the published $68.3\%$ and $95.4\%$ contours for PSR J0740+6620~\cite{Salmi2024J0740Data}. Teal solid and dotted curves show the $68\%$ and $95\%$ highest-density contours reconstructed from the weighted PSR J0437-4715 samples~\cite{Miller2026J0437Data}.}
\label{fig:observationmr}
\end{figure*}

\subsection{Likelihoods from public posterior samples}
We reweight the public posteriors within the adopted prior bounds, retaining each event's original marginal mass prior~\cite{ThraneTalbot2019}. For NICER source $i$, we divide the joint mass--radius posterior by the adopted conditional radius prior. Integrating this ratio along the model curve gives
\begin{equation}
\mathcal L_i(\theta)\propto
\int\frac{p_{0,i}[M,R(M;\theta)\mid d_i]}
{p_{0,i}[R(M;\theta)\mid M]}\,dM,
\label{eq:nicerlike}
\end{equation}
where $p_{0,i}(M,R\mid d_i)$ denotes the published posterior given data $d_i$, and $p_{0,i}(R\mid M)$ denotes the adopted radius prior at fixed mass. For PSR J0740+6620, we use the published constraints to reconstruct the allowed radius interval at each mass and assume a uniform conditional prior within it. The reconstruction and the density estimates are described in Appendix~\ref{app:conditionalprior}. The quoted numerical errors exclude uncertainty in the reconstructed prior.

For GW event $j$, we replace the original conditional tidal prior with the product of the two component proxy densities from Eq.~\eqref{eq:proxy}. Averaging the ratio of the new density to the original prior over the binary posterior gives
\begin{equation}
\mathcal L_j(\theta)\propto
\sum_s w_s
\frac{\displaystyle\prod_{A=1}^{2}
p_{\rm prox}[\Lambda_{A,s}\mid M_{A,s},R(M_{A,s};\theta)]}
{p_{0,j}(\Lambda_{1,s},\Lambda_{2,s}\mid M_{1,s},M_{2,s})},
\label{eq:gwlike}
\end{equation}
where $w_s$ is the normalized weight of binary sample $s$ and $A$ labels its two stars. The denominator is the original joint tidal prior at fixed component masses. In both events the component tidal priors are independent and uniform. For GW170817 both range from $0$ to $5000$. For GW190425 the ranges are $[0,5000]$ for the heavier star and $[0,10000]$ for the lighter star~\cite{Wang2026GW170817,WangGWPosteriorData,Abbott2020GW190425,Abbott2024GWTC21,GW190425Data2022}.

NICER masses outside the model's computed range contribute zero to Eq.~\eqref{eq:nicerlike}. A GW sample contributes only if the model covers both component masses. We do not renormalize the mass integral or the sample weights after this restriction, so unsupported masses reduce the event likelihood.

For independent events the joint posterior is
\begin{equation}
p(\theta\mid d)\propto p(\theta)
\prod_{i\in\mathrm{NICER}}\mathcal L_i(\theta)
\prod_{j\in\mathrm{GW}}\mathcal L_j(\theta),
\label{eq:jointposterior}
\end{equation}
Here $d$ denotes the combined data. For NICER alone, GW alone, and NICER+GW we use the same prior $p(\theta)$, uniform on $\lsym\in[10,200]$ MeV and $\mu_1\in[-1,1]$ subject to the EOS and stellar-solution validity conditions. For the GR comparison we fix $\mu_1=0$ and use the same four sources. We compare the resulting parameter posteriors for the two models.

\subsection{Parameter constraints and sensitivity to the observational assumptions}
In FIG.~\ref{fig:observationcorner}, the posteriors from NICER, GW, and all four sources within the EOS family of TABLE~\ref{tab:eos} are compared. The two pulsars alone give a median $\lsym$ of $125$ MeV. Adding GW170817 lowers it to $73$ MeV, whereas adding GW190425 alone lowers it to $114$ MeV (see TABLE~\ref{tab:constraints}). GW170817 therefore accounts for most of the shift toward smaller $\lsym$ and most of the reduction in the upper endpoint of the central $68\%$ interval.

\begin{table}[tb]
\caption{Marginal medians and central $68\%$ credible intervals for the fixed EOS inputs of Table~\ref{tab:eos}. NICER denotes both pulsars and GW denotes both mergers. Only the last row fixes $\mu_1=0$.}
\label{tab:constraints}
\small
\setlength{\tabcolsep}{2pt}
\begin{ruledtabular}
\begin{tabular}{lcc}
Data & $\lsym$ (MeV) & $\mu_1$\\
NICER & $125\ [63,176]$ & $0.41\ [0.04,0.81]$\\
GW & $48\ [20,108]$ & $0.38\ [-0.14,0.81]$\\
NICER + GW190425 & $114\ [57,169]$ & $0.42\ [0.05,0.81]$\\
NICER + GW170817 & $73\ [28,135]$ & $0.47\ [0.09,0.83]$\\
NICER + GW & $69\ [28,127]$ & $0.47\ [0.10,0.83]$\\
NICER + GW (GR) & $73\ [28,128]$ & $0$ (fixed)\\
\end{tabular}
\end{ruledtabular}
\end{table}

For the same four sources, allowing $\mu_1$ to vary lowers the median $\lsym$ from $73$ to $69$ MeV and leaves its central $68\%$ interval nearly unchanged (see FIG.~\ref{fig:observationL}). The central $95\%$ interval for $\mu_1$ is $[-0.09,0.97]$ and includes the GR value. Its upper endpoint lies close to the prior bound $\mu_1=1$. The posterior correlation between $\lsym$ and $\mu_1$ is $-0.007$. The radius fits in Sec.~\ref{sec:degeneracy} compare each candidate with a specified GR sequence at equally weighted masses. The posterior instead weights each model by its prior and the four source likelihoods.

We test how the median shift depends on the allowed range of the gravity coefficient. Narrowing the uniform $\mu_1$ prior from $[-1,1]$ to $[-0.5,0.5]$ and $[-0.25,0.25]$ reduces the shift relative to GR from $-4.06$ to $-3.71$ and $-0.76$ MeV, respectively. The corresponding post-TOV $68\%$ intervals are $[28,127]$, $[26,127]$, and $[27,128]$ MeV.

Restricting the analysis to models in which the internal RMS correction at $M=1.4M_\odot$ is at most one tenth of the GR term leaves a median difference of $0.25$ MeV (see Appendix~\ref{app:finiteamplitude}).

We next vary the half-width of the compactness residual in Eq.~\eqref{eq:proxy}, using the same width in GR and post-TOV. Halving or doubling the baseline width gives post-TOV $68\%$ intervals of $[27,122]$ and $[31,128]$ MeV, respectively. At half, baseline, and double width, the post-TOV median minus the GR median is $-7.37$, $-4.06$, and $-1.66$ MeV. The corresponding standard deviations from $24$ paired resamplings of the GW samples are $1.80$, $0.71$, and $0.38$ MeV. At the baseline width, allowing residual correlations of $-1/3$ or $+1/3$ between the two stars also leaves the GR and post-TOV credible intervals largely overlapping (Appendix~\ref{app:observationchecks}).

Omitting J0740+6620 removes the dependence on its reconstructed conditional radius prior along with the information from that source. J0437$-$4715 and the two GW events give a post-TOV median of $68$ MeV and a central $68\%$ interval of $[25,131]$ MeV, compared with $65$ MeV and $[22,127]$ MeV in GR. The intervals still overlap, but the sign of the median difference changes. This omission tests sensitivity to the inclusion of J0740+6620 without isolating uncertainty in its prior.

The corresponding mass--radius predictions for all four sources are given in FIG.~\ref{fig:observationmr}. At $M=1.4M_\odot$, the median radius is $12.61$ km in GR and $12.37$ km in post-TOV. The central $68\%$ radius intervals, $[11.85,13.32]$ km and $[11.63,13.19]$ km, overlap substantially.

\FloatBarrier

\section{Conclusions and discussions}
\label{sec:discussion}
Near the $\lsym=40$ MeV GR reference, a single post-TOV coefficient can partly compensate for changes in the EOS over $1$--$2M_\odot$. The amount of compensation depends on which mass is held fixed. At fixed enclosed mass, fitting $\mu_1$ reduces the RMS radius difference by about $64\%$ for both $\lsym=30$ and $50$ MeV. At fixed asymptotic mass the reduction is $17\%$ and $75\%$, respectively. At the lower slope, the more negative coefficient favored by the enclosed-mass fit gives a sequence that ends at $M=1.84M_\odot$. The full comparison interval in $M$ therefore calls for a different fit.

Within the EOS families and with the adopted crust--core matching, the four sources constrain $\lsym$ to nearly the same range in GR and in the single-$\mu_1$ model. The central $68\%$ credible intervals are $[28,128]$ and $[28,127]$ MeV, respectively. The medians are $73$ and $69$ MeV, but their difference depends on the prior for $\mu_1$, the tidal proxy, and the sources included. The $95\%$ interval for $\mu_1$ includes zero. Of the two mergers, GW170817 accounts for most of the narrowing relative to NICER alone. The weak posterior correlation between $\lsym$ and $\mu_1$ does not by itself resolve the radius degeneracy. The posterior weights sequences by the four observed sources, whereas the radius fits compare them with one specified GR sequence.

We selected $\mu_1$ by comparing the relative sizes of the corrections in the stellar structure equations. How the inferred range of $\lsym$ would change if other post-TOV coefficients were also allowed to vary remains an open question. The observational conclusions apply to the specified EOS family, with the other nuclear inputs and high-density sound speed fixed. The published pulsar and GW analyses assume GR, and the adopted single-coefficient model supplies no tidal perturbation equations. Extending the comparison to a specific theory of gravity would also require its predicted tidal response and its effects on the observed signals.

\begin{acknowledgments}
We thank Xing-Hua Jin for helpful suggestions. We thank the CompOSE team for assistance with account registration and the EOS contributors for providing the equation-of-state tables used in this work. We acknowledge the developers and maintainers of JESTER and X-PSI for making their software publicly available.
\end{acknowledgments}


\appendix
\section{Numerical methods and verification}
\label{app:numerics}
\subsection{Calibration checks}
\label{app:calibrationchecks}
Under numerical refinement, the largest relative changes in the calibrated amplitudes are $9.94\times10^{-7}$ on fixed GR profiles and $4.61\times10^{-7}$ on modified profiles. The largest relative residuals are $1.15\times10^{-8}$ in the target mass and $4.23\times10^{-8}$ in the target strength. Independent integration of the $300$ single-coefficient solutions reproduces their masses, radii, and strengths to within $10^{-6}$ in relative terms. For the $21$ equal-coefficient solutions, the relative differences in mass, radius, and each $Q_j$ are below $10^{-7}$.

We also compare the radius responses to $\mu_1$ and $\mu_2$ in the $\lsym=40$ MeV EOS of Sec.~\ref{sec:eosfamily}. We calibrate each coefficient separately, for each sign, to $\eta=10^{-3}$ at $\ms=1.4M_\odot$ and keep it fixed along the resulting sequence. At each asymptotic mass $M$ we take the radius difference between the positive- and negative-coefficient sequences and divide by $2\eta$. The RMS of this quantity over the $161$ comparison masses in Eq.~\eqref{eq:distance} is $2.63$ km for $\mu_1$ and $4.30$ km for $\mu_2$. Thus the coefficient with the largest correction per unit amplitude need not give the largest radius change at equal correction strength.

\subsection{EOS construction and enthalpy integration}
\label{app:eosimplementation}
The metamodel uses $b(\delta)=17+25\delta^2$~\cite{Somasundaram2021}, where $\delta=(n_n-n_p)/n$. We retain the full free Fermi-gas kinetic term and include the nucleon rest masses and ideal relativistic electrons in the energy minimization. For comparison with the quadratic convention, let $S_2(n)$ be the coefficient of $\delta^2$ in the expansion about symmetric matter. Our inputs in Eq.~\eqref{eq:symmetrydefinition} give $S_2(\nsat)=31.296$ MeV and $3\nsat S_2'(\nsat)=\lsym-1.408~\mathrm{MeV}$.

We preserve the low-density sampling of the original 152-row Douchin--Haensel crust table. The cubic Hermite interpolation of $c_s^2$ matches both endpoint values and their density derivatives at the crust and metamodel boundaries. Starting from the lowest crust state, we set the baryon chemical potential to $\mu_{\rm B}=(\varepsilon+p)/n$ there and integrate $d\ln\mu_{\rm B}/d\ln n=c_s^2$, $dp/dn=c_s^2\mu_{\rm B}$, and $d\varepsilon/dn=\mu_{\rm B}$ to construct the EOS. We use shape-preserving cubic Hermite interpolation of $\ln\varepsilon$ and $\ln n$ in $\ln p$. We retain every original pressure knot when constructing the $35\,653$-point tables. For the dimensionless enthalpy we use
\begin{equation}
 h(p)=\int_{p_{\rm cut}}^{p}\frac{dp'}{\varepsilon(p')+p'} .
 \label{eq:enthalpy}
\end{equation}
We apply six-point Gauss--Legendre quadrature within each interpolation piece, using this interpolant for $\varepsilon(p)$. During stellar integration we interpolate the tabulated thermodynamic variables linearly in logarithmic coordinates. We assign $h(p_{\rm cut})=10^{-30}$ for logarithmic interpolation. As an independent check, direct integration of the GR structure equations in $\ln p$ reproduces the enthalpy-coordinate radii at $1.4M_\odot$ within $0.002$ m for $\lsym=30$, $40$, and $50$ MeV.

The matched sound speed specifies $dp/d\varepsilon$ at the core boundary. The values of $p$ and $\mu_{\rm B}$ reached there also depend on the crust, where the integration starts. In TABLE~\ref{tab:eospressure}, the resulting pressure at $\nsat$ is compared with that of uniform neutron--proton--electron matter. For the latter we evaluate $p=n(\partial\varepsilon/\partial n)_{Y_p}-\varepsilon$ at the equilibrium proton fraction $Y_p$. From $\lsym=30$ to $50$ MeV the pressure at saturation rises by $0.877~\mathrm{MeV\,fm}^{-3}$ in the uniform-matter calculation and by $0.538~\mathrm{MeV\,fm}^{-3}$ in the stellar EOS. Assessing the effect on radii requires a comparison of complete EOSs with different crust--core matches~\cite{Fortin2016Crust}.

\begin{table}[tb]
\caption{Pressure at $n=\nsat$ from the uniform-matter functional ($p_{\rm func}$) and the stellar EOS ($p_{\rm star}$), in $\mathrm{MeV\,fm}^{-3}$. The fourth column gives $(p_{\rm star}-p_{\rm func})/p_{\rm func}$ in percent. The last gives the difference in $\mu_{\rm B}=(\varepsilon+p)/n$, stellar minus functional.}
\label{tab:eospressure}
\begin{ruledtabular}
\begin{tabular}{rccrr}
$\lsym$ (MeV) & $p_{\rm func}$ & $p_{\rm star}$ & $\Delta p/p$ (\%) & $\Delta\mu_{\rm B}$ (MeV)\\
\hline
$30$ & $1.521$ & $1.936$ & $+27.3$ & $+2.51$\\
$40$ & $1.959$ & $2.209$ & $+12.8$ & $+1.91$\\
$50$ & $2.397$ & $2.474$ & $+3.2$  & $+1.26$\\
\end{tabular}
\end{ruledtabular}
\end{table}

\subsection{Stellar sequences and radius fits}
\label{sec:verification}
We integrate the post-TOV equations in logarithmic enthalpy~\cite{Lindblom1992} using Tsit5~\cite{Tsitouras2011} in double precision. The relative and absolute tolerances are $10^{-8}$ and $10^{-10}$. Starting $2\times10^{-5}h_{\rm c}$ below the central enthalpy $h_{\rm c}$, we integrate to the lowest tabulated pressure. For each sequence we use $512$ logarithmically spaced central pressures, starting at the first tabulated pressure with $n\geq\nsat$ and ending at the highest tabulated pressure. We refine the sampling near the enclosed-mass peak and any interior peak in the mapped mass. Accepted interiors have finite structure variables, positive density and the condition $1-2m/r>0$ is satisfied.

For comparisons at fixed $M$ we retain the first connected, exterior-valid run that reaches $1M_\odot$. A decrease of more than $10^{-3}M_\odot$ from the running maximum terminates the branch. Small nonincreasing steps are omitted before interpolation. For the fixed-$\ms$ comparison we split at each nonincreasing step and require a unique segment containing $1M_\odot$. The former selection tests the exterior matching condition, whereas the latter requires only an interior solution. Each radius fit requires coverage of $[1,2]M_\odot$ in its own mass coordinate. The observational likelihood uses all masses supported by the selected $M$ branch and does not impose this two-solar-mass requirement.

\begin{figure*}[t]
\centering
\includegraphics[width=\textwidth]{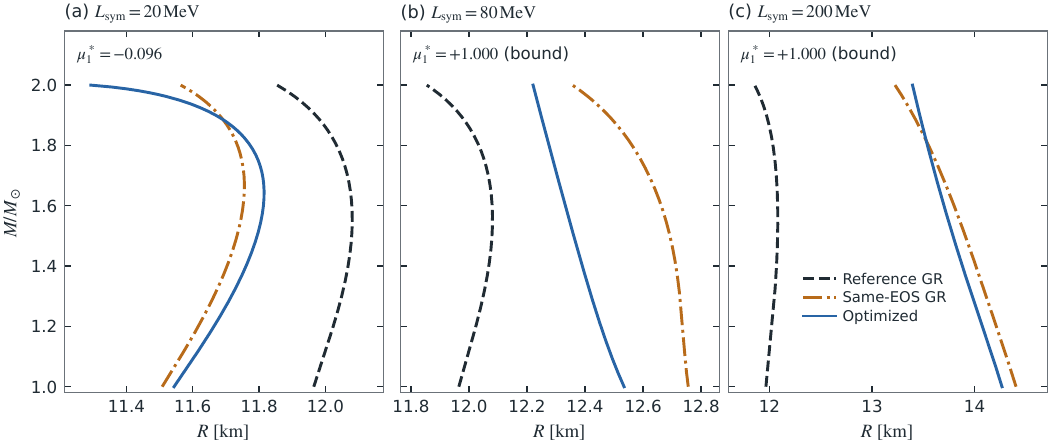}
\caption{Mass--radius curves at fixed asymptotic mass for three fits in Table~\ref{tab:results}. Each panel compares the $L_0=40$ MeV GR reference, the candidate EOS in GR, and its fitted sequence. At $80$ and $200$ MeV the fitted coefficient reaches $\mstar=1$. Note that the radius scales differ among the panels.}
\label{fig:examples}
\end{figure*}

For each mass coordinate, we refine the coefficient search near candidate minima and near the boundaries where a sequence ceases to cover the mass window. To check the four near-reference fits, we hold their fitted coefficients fixed and increase the sampling to $1024$ central pressures. We tighten the tolerances to $2\times10^{-10}$ and $2\times10^{-12}$ and reduce the initial offset to $2\times10^{-6}h_{\rm c}$. The largest change in $\dr$ is $4.9\times10^{-7}$ km, and the largest radius difference at a single comparison mass is $0.042$ m. 

\subsection{Conditional radius priors and density transformation}
\label{app:conditionalprior}
The NICER likelihood in Eq.~\eqref{eq:nicerlike} uses the conditional radius prior for each source $i$. At fixed mass $M$, we use the density
\begin{equation}
p_{0,i}(R\mid M)=\frac{\mathbf{1}_{[R_{-,i}(M),R_{+,i}(M)]}(R)}
{R_{+,i}(M)-R_{-,i}(M)}.
\label{eq:conditionalradiusprior}
\end{equation}
Here $R_{-,i}(M)$ and $R_{+,i}(M)$ are the lower and upper radius bounds, and the indicator equals one inside the interval and zero outside. For J0437$-$4715, this density follows from the published inverse-compactness prior, with $R_-=3.2GM/c^2$ and $R_+=8GM/c^2$~\cite{Miller2026J0437}.

For J0740+6620, we reconstruct the allowed radius interval at each mass from the published constraints and assume a uniform prior in radius over that interval. For $M\in[1,3]M_\odot$, we start from $[3GM_\odot/c^2,16~\mathrm{km}]$ and impose the polar-radius condition $R_{\rm p}>3GM/c^2$ and the effective-surface-gravity range $13.7\leq\log_{10}[g/(\mathrm{cm\,s^{-2}})]\leq15$ at the pole and equator~\cite{Salmi2024J0740}. We evaluate the radius and effective gravity with X-PSI v1.2.1~\cite{Riley2023XPSI} at the adopted rotation period of $2.8857$~ms. The surface-shape constraint is inactive over the tested domain. The allowed radii form a single interval with $R_+=16$ km. At $M=2M_\odot$, for example, $R_-\simeq8.91003$ km. These restrictions can change the marginal mass prior from its untruncated form.

For density estimation, we map each mass-dependent radius interval to the same unit interval. Within the conditional-prior support,
\begin{equation}
u=F_{0,i}(R\mid M)
=\frac{R-R_{-,i}(M)}{R_{+,i}(M)-R_{-,i}(M)}.
\label{eq:priorcoordinate}
\end{equation}
Here $F_{0,i}(R\mid M)$ is the cumulative distribution associated with Eq.~\eqref{eq:conditionalradiusprior}. The Jacobian $\partial u/\partial R=p_{0,i}(R\mid M)$ gives the posterior density with respect to $dM\,du$
\begin{equation}
\widetilde p_{0,i}(M,u\mid d_i)
=\left.\frac{p_{0,i}(M,R\mid d_i)}{p_{0,i}(R\mid M)}
\right|_{R=F_{0,i}^{-1}(u\mid M)}.
\label{eq:transformedposterior}
\end{equation}
The transformation allows for dependence between mass and radius. We estimate the density of the weighted $(M,u)$ samples with Gaussian kernels, Scott's bandwidth rule~\cite{Scott1992}, and reflection at both boundaries of the unit interval~\cite{Schuster1985}. We evaluate this density along $u=F_{0,i}[R(M;\theta)\mid M]$ and integrate over $M$ in Eq.~\eqref{eq:nicerlike}. A predicted radius outside the allowed interval contributes zero at that mass. The likelihood retains the induced marginal mass prior, so we apply no additional radio mass prior.

We check the J0740 bounds and their monotonicity on $1001$ mass points with $2049$ radius points per mass. All $4\,109\,423$ samples from the two pulsars lie within their respective bounds. Whether the full J0740 nuisance-parameter prior induces the assumed uniform density at fixed mass remains unverified. The numerical error estimates exclude uncertainty in this reconstruction.

\subsection{Numerical checks of the four-source posterior}
\label{app:observationchecks}
The observational grid contains $77\times41=3157$ sequences with spacings of $2.5$ MeV in $\lsym$ and $0.05$ in $\mu_1$. All $1\,691\,844$ interior integrations completed, and each sequence has a branch usable in the observational likelihood. An endpoint set by the upper density limit or exterior matching restricts the mass coverage. It need not be a maximum stable mass.

We integrate the piecewise-linear marginal densities and invert the resulting quadratic cumulative distribution on each interval. We check the $2.5$, $16$, $50$, $84$, and $97.5$ percentiles. Doubling the parameter-grid spacings changes these percentiles by less than $0.05$ MeV in $\lsym$ and $0.009$ in $\mu_1$, and by less than $0.065$ MeV in the GR value of $\lsym$. Halving the NICER mass-integration spacing from $5\times10^{-4}M_\odot$ changes all $\lsym$ percentiles by less than $3.4\times10^{-4}$ MeV and all $\mu_1$ percentiles by less than $1.1\times10^{-5}$. Scaling both NICER kernel bandwidths by $0.8$ and $1.2$ changes the $\lsym$ percentiles by at most $0.123$ MeV in post-TOV and $0.156$ MeV in GR. The largest change in $\mu_1$ is below $2.0\times10^{-4}$.

In each of $24$ normalized Poisson resamplings of the GW posteriors, we keep both component masses and tidal deformabilities together and use the same resampling weights in GR and post-TOV. At the baseline proxy width, the standard deviations of the four-source $\lsym$ medians are $2.46$ MeV in post-TOV and $2.14$ MeV in GR. We take the standard deviation of the paired differences to quantify the effect of finite GW samples on the median shift at each width.

To test the assumption that the two compactness residuals are independent, we multiply the product of the proxy densities in Eq.~\eqref{eq:gwlike} by $1+\kappa t_1t_2$. We require $|\kappa|\leq 1$ to keep the joint density nonnegative. The scaled residual for star $A$ is
\begin{equation*}
t_A=\frac{C_{\rm fit}(\Lambda_A)/C_A-1}{0.065}, \qquad A=1,2 .
\end{equation*}
Here $C_A$ is the compactness of star $A$. Each $t_A$ retains its uniform distribution on $[-1,1]$, and the residual correlation is $\kappa/3$. At $\kappa=0$ we recover the baseline event likelihoods. For a test correlation of $-1/3$, the four-source $68\%$ intervals for $\lsym$ are $[27.2,126.2]$ MeV in post-TOV and $[27.0,126.4]$ MeV in GR. For $+1/3$ they are $[29.0,128.1]$ and $[28.2,129.3]$ MeV, respectively. The corresponding post-TOV minus GR median differences are $-4.95$ and $-3.13$ MeV.

\par\vfil
\subsection{Correction strength at fixed observed mass and a GR tidal check}
\label{app:finiteamplitude}

For each four-source parameter-grid point we compute the radial RMS ratio of the $\mu_1$ term to the GR term in the mass equation at $M=1.4$ and $2.0M_\odot$, wherever the selected branch reaches that mass. We integrate the squares of the two terms with respect to radius alongside the structure equations and adjust the central pressure to match the specified asymptotic mass. We also evaluate the fractional change $(M-\ms)/\ms$ from the exterior mass relation. At each mass we weight the supported models by the four-source posterior and renormalize. The supported fractions of the posterior are $100\%$ and $99.5\%$, respectively. The median internal RMS ratios are $0.069$ and $0.113$, with central $68\%$ intervals of $[0.017,0.112]$ and $[0.034,0.172]$. The median fractional mass changes are $7.9\%$ and $16.1\%$. Both quantities measure effects of the retained term. An estimate of the omitted higher-order terms requires additional assumptions.

To test sensitivity to larger corrections, we restrict the inference to models whose internal RMS ratio at $M=1.4M_\odot$ is at most $0.10$. Every sequence reaches this mass. The restriction retains $73.8\%$ of the four-source posterior weight. The $\lsym$ median becomes $73$ MeV and the central $68\%$ credible interval becomes $[28,131]$ MeV. The median differs from the GR result by $0.25$ MeV, compared with $4.06$ MeV before the restriction. The $10\%$ threshold is a sensitivity test of the retained correction at one mass.

We check the tidal proxy in GR by solving the quadrupolar tidal perturbation equation for $25$ stars in the same EOS family. The tests cover $\lsym=10$, $30$, $40$, $50$, $100$, and $200$ MeV and masses from $1.0$ to $2.08M_\odot$. Using the calculated tidal deformability in the compactness fit gives fractional residuals $(C_{\rm fit}-C)/C$ between $-1.93\%$ and $+3.09\%$. All lie within the adopted $\pm6.5\%$ range. For $\mu_1\ne0$ the tidal response is still supplied by the proxy.

\bibliography{refs}
\end{document}